\documentclass[conference]{IEEEtran}
\IEEEoverridecommandlockouts
\usepackage{cite}
\usepackage{amsmath,amssymb,amsfonts}
\usepackage{algorithmic}
\usepackage{graphicx}
\usepackage{textcomp}
\usepackage{xcolor}

\graphicspath{{pdf/}{jpeg/}{png/}{tapia_figures/}{figures/}}
\usepackage{graphicx}
\usepackage{algorithmic}
\usepackage{multirow}
\usepackage{listings}
\usepackage[caption=false,font=footnotesize]{subfig}
\usepackage[bookmarks=false]{hyperref}
\hypersetup{
     filecolor=black,
     allbordercolors = white
}

\usepackage{relsize}
\usepackage{comment}

\usepackage{stfloats}
\usepackage{balance}
\usepackage{soul}
\usepackage[normalem]{ulem}
\usepackage{cancel}
\usepackage{xcolor}
\usepackage{tabularx}

\usepackage{balance}

\newcommand{\oma}[1]{{\color{blue}{\textbf{SHINA:} #1}}}

\usepackage{soul}  
    
\begin{document}

\title{Using Delay Tolerant Networks as a Backbone for Low-cost Smart Cities\\}

\author{
    \IEEEauthorblockN{Oluwashina Madamori\IEEEauthorrefmark{1}, Esther Max-Onakpoya\IEEEauthorrefmark{1}, Christan Grant\IEEEauthorrefmark{2}, Corey E. Baker\IEEEauthorrefmark{1}}
    \IEEEauthorblockA{
        \IEEEauthorrefmark{1}Department of Computer Science, 
        University of Kentucky, Lexington, KY USA
    }
    \IEEEauthorblockA{
        \IEEEauthorrefmark{2}Department of Computer Science, 
        University of Oklahoma, Norman, OK USA \\
        shina@uky.edu, esther.max05@uky.edu, cgrant@ou.edu, baker@cs.uky.edu 
    }
}

\maketitle

\begin{abstract}
Rapid urbanization burdens city infrastructure and creates the need for local governments to maximize the usage of resources to serve its citizens. Smart city projects aim to alleviate the urbanization problem by deploying a vast amount of Internet-of-things (IoT) devices to monitor and manage environmental conditions and infrastructure. However, smart city projects can be extremely expensive to deploy and manage. A significant portion of the expense is a result of providing Internet connectivity via 5G or WiFi to IoT devices. This paper proposes the use of delay tolerant networks (DTNs) as a backbone for smart city communication; enabling developing communities to become smart cities at a fraction of the cost. A model is introduced to aid policy makers in designing and evaluating the expected performance of such networks. Preliminary results are presented based on a public transit network data-set from Chapel Hill, North Carolina. Finally, innovative ways of improving network performance in a low-cost smart city is discussed.  
\end{abstract}


\section{Introduction}
Urbanization, which refers to the migration of people from rural areas to urban communities, has increased rapidly in the past few decades 
%
%
placing an extra burden on the infrastructure in urban areas~\cite{UnitedNations2018}. 
Policy makers are thus tasked with finding more effective methods for managing limited public resources to meet the needs of its citizens. Smart city initiatives have been proposed as a way to alleviate the challenges arising from urbanization. Smart cities projects involve the deployment of a vast number of Internet-of-things (IoT) devices across communities to monitor and manage environmental conditions and infrastructure. 
However, transforming a city into a smart city has proven to be very expensive, for example, cities such as San Diego, 
New Orleans, London, and Songdo have either proposed or invested in smart city projects that cost between  \$30 Million and \$40 Billion~\cite{SanDiegoSmartCity, Technology, SmartLondonBoard2013, Insider}.
%
As a result, financing smart city projects is a major challenge that limits its implementation~\cite{Deloitte2018,SmartCitiesWorld}. Therefore, finding ways to significantly reduce the cost of transforming traditional cities into smart cities is critical. 

Many smart city designs require IoT devices to be connected to the Internet in order to retrieve the data generated by the devices. 
Unfortunately, significant costs are incurred when deploying sensors equipped with 5G or WiFi connectivity due to data subscription fees~\cite{paradells2014,max2019opportunistic}.
This work proposes a low-cost alternative to cellular and always-connected wireless connectivity using a delay tolerant network (DTN) that leverages the pre-existing mobility of public transportation and pedestrians to create opportunistic communication networks for delivering IoT data to the cloud. 

\begin{figure}[htbp]
	\centering
	\includegraphics[width=0.48\textwidth,height=2.0in]{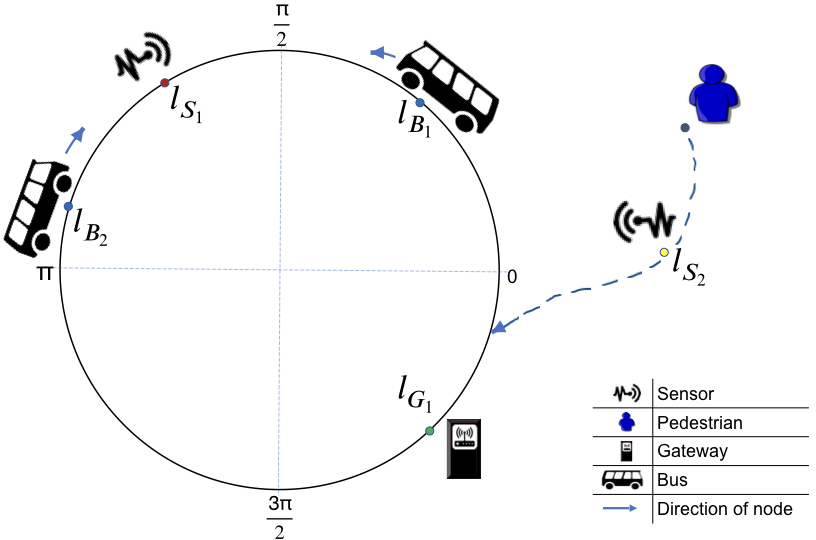}
    \caption{Opportunistic communication for low-cost smart cities}
	\label{flow_diagram}
\end{figure}

The goal is to provide city planners with a viable and cheaper alternative to Internet connectivity for IoT devices in smart cities, by eliminating or reducing the associated Internet subscription fees.
Previous research~\cite{morgenroth2015performance} has shown that the two main limitations of DTNs compared to cellular networks are: (i) low and unpredictable probability of delivery, and (ii) high and unpredictable network latency. In order for DTNs to serve as a backbone for smart city communication, the aforementioned limitations need to be addressed so that developers can ensure that such opportunistic networks meets quality of service (QoS) requirements~\cite{baker2017vivo}.

To demonstrate that DTNs can serve as a backbone for smart city communication, a network model is developed that characterizes the entities participating in the DTN and how they interact with one another.
The model can also help policy makers understand how changes in their transit system with respect to transit scheduling, gateway deployment, sensor placement, ridership levels, and number of participating citizens can impact performance in their low-cost smart city. 
The rest of the paper is structured as follows: Section \ref{sec:dataset} describes the tools used in this work to capture data-sets from real-world public transit systems. Section \ref{sec:modeling} describes the proposed model. Section \ref{sec:results} provides preliminary results. Finally, Section \ref{sec:future} describes the next steps in this work and the need for incentivization strategies that can be  used  to  maximize  delivery probability and minimize latency.
%
%

\section{Understanding City Public Transportation} \label{sec:dataset}
Many cities currently provide highly reliable information on the Internet about their public transit network. Popular sources include \textit{OpenMobilityOrg} which contains General Transit Feed Specification (GTFS) data~\cite{OpenMobilityOrgOpenMobilityDataWorld} and \textit{NextBus}~\cite{CUBIC}. These data includes static information that specify bus routes, stops, and operating schedules. 
In addition, OpenMobilityOrg and NextBus provide real-time information about GPS location and expected arrival times of buses within various city transit networks.
%
To validate the model presented in this work, the \textit{NextBus} API was used to obtain data about the bus routes in the city of Chapel Hill, North Carolina. 
The data retreived contained descriptions of bus routes, trip directions, and stops per route for all buses during operation during the month of July, 2018. 
The characteristics of Chapel Hill public transportation are listed in Table~\ref{route_stats}. 
%

\begin{table}[htbp]
\centering
\caption{Chapel Hill Public Bus Characteristics}
\label{route_stats}
\begin{tabular}{|l|l|}
\hline
Numbers of routes & 32 \\ \hline
Distance of routes & \begin{tabular}[c]{@{}l@{}}$\mu=15.70 $ km, $\sigma^{2} = 7.32$ km\end{tabular} \\ \hline
Stops per route & \begin{tabular}[c]{@{}l@{}}$\mu=72.6$, $\sigma^{2} = 52.44$\end{tabular} \\ \hline
Buses per route & \begin{tabular}[c]{@{}l@{}}$\mu=1.79$, $\sigma^{2} =1.00$\end{tabular} \\ \hline
\end{tabular}
\end{table}

\section{Modeling Low-cost Smart Cities}\label{sec:modeling}

A smart city consists of many entities. The entities along with assumptions about their characteristics are as follows:   
\begin{itemize}

\item \textbf{Bus routes} - are taken by buses and due to the recurring nature of buses visiting the same stop, can be represented by circles, each with a circumference corresponding to the total distance for that route. In addition, the locations of all buses, bus stops, gateways, and on-route sensors are restricted to points on a circle representing a bus route. 
		 	 	 							
\item \textbf{Buses} - move along predefined routes on a fixed schedule, hence, the specific geographic position of any bus can be calculated at any time. Buses are also expected to move at a constant average velocity (including stops) throughout a route trip, thus assigning every bus a fixed round-trip time. Buses have buffer/queue sizes of infinity and do not use any type of drop-policy for stored data. 

\item \textbf{Pedestrians} - people who sign-up to join the low-cost smart city using their smartphones. Data packets from sensors are forwarded to their smartphones using Bluetooth 5 when they come within transmission range of a sensor. The data is stored on the phone until the pedestrian comes within transmission range of a bus or gateway and forwards the data to that node.

\item \textbf{Sensors} - 
broadly classified into two categories based on their location: on-route and off-route sensors. \textbf{On-route sensors} are those located at stop-lights, street-lights, and bus stops; and are within transmission range of buses travelling on a bus route. Any sensor that is not an ``on-route sensor'' is classified as an \textbf{off-route} sensor. Off-route sensors have an additional parameter associated with them called the ``pedestrian arrival rate,'' which specifies the likelihood of a participating pedestrian coming in contact with the sensor.
Although pedestrians are not essential for retrieving on-route sensor data, they are vital in the retrieval of off-route sensor data. In addition, individual data packets generated by sensors are considered to be small, allowing sensor buffer/queue size to be infinity and no drop policy needed. 
					
\item \textbf{Gateways} - stationary, always-on, always-connected devices that forward data directly to the cloud. Not all bus stops are gateways, but rather gateways are placed at select bus stops. Gateways act as the final destination for all data generated by sensors. 
					
\end{itemize}

\subsection{Assumptions}\label{sec_assumptions}

The model makes several assumptions about the DTN within a smart city. Each bus route has at least one gateway, and buses operate round the clock everyday. Furthermore, a connection is always established between the sensor and bus (or pedestrian) whenever a bus (or pedestrian) passes a sensor, and all data at the sensor is transferred to the bus (or pedestrian). In addition, there is no routing from pedestrian to pedestrian or from bus to bus.

As the model is further developed some of the aforementioned assumptions will be relaxed. Figure \ref{flow_diagram} summarizes the interactions between entities in the model.




\subsection{Estimating delivery latency}
\textbf{On-route sensors:} \label{on-route delay}
The expected delivery latency $T_{D\text{-on}}$ for any data produced at a sensor at an arbitrary time $t$ is represented by:

\begin{equation}\label{}
T_{D\text{-on}} = T_{SB} + T_{BG} 
\end{equation}

\begin{equation}\label{}
 T_{SB} = \frac{\textit{distance from bus location to sensor at  }t}{\textit{average bus velocity}}
\end{equation}

\begin{equation}\label{}
T_{BG} = \frac{\textit{distance from bus location to gateway at } t+T_{SB}} {\textit{average bus velocity}}
\end{equation}

\noindent and the upper bound  $\max(T_{D\text{-on}})$ for delivery latency will be: 

\begin{equation}\label{}
 \max(T_{D\text{-on}}) \approx \frac{ 2 \times \textit{circumference of route}}{\textit{average bus velocity}}
\end{equation}

\textbf{Off-route sensors:}
For off-route sensors, there are two additional stages between when the sensor data is generated and when it is delivered at the gateway.

\begin{equation}\label{}
 T_{D\text{-off}} = E\big[ T_{SP} \big] + E\big[ T_{PB} \big] + T_{SB} + T_{BG} 
\end{equation}

\noindent Where $E\big[ T_{SP} \big]$ is the expected time it takes a participating pedestrian to encounter the sensor after the data is generated in time $t_0$, 
and $E\big[ T_{PB} \big]$ is the expected duration it takes for that pedestrian to board (or encounter) a bus. 
%
%
Also, the upper bound  $\max (T_{D\text{-off}})$ for delivery latency will be:
\begin{equation}\label{eqn_test}
 \max(T_{D\text{-off}}) = \max(T_{SP})+\max(T_{PB}) + \max(T_{D\text{-on}})
\end{equation}

\section{Preliminary Results}\label{sec:results}
Using the tools described in Section~\ref{sec:dataset} to generate a data-set for Chapel Hill, North Carolina along with a 2016 Chapel Hill transit passenger survey~\cite{Institute2016}, a light-weight simulator\footnote{The code for the simulator is available on GitHub - \url{https://github.com/netreconlab/smartcomp19}} was developed to find the expected performance of transforming Chapel Hill into a low-cost smart city.
The simulator is unique because it uses basic input parameters such as the number or routes, number of buses and stops, etc. for Chapel Hill (or any city) and returns an upper-bound of network performance. 
The input parameters for the simulation are listed in Table~\ref{sim_parameters}.

\setlength{\belowcaptionskip}{0pt}

\begin{table}[htbp]
\centering
\caption{Parameters used in the Simulation}
\label{sim_parameters}
\begin{tabular}{|l|l|}
\hline
Simulation seeds & 0:1:99\\ \hline
Simulation duration & 48 hours\\ \hline
Number of routes & 32 \\ \hline
On-route sensors per route & 2 to 8 \\ \hline
Gateways per route ($G$) & 1 to 2 \\ \hline
Number of bus stops per route ($S$) & \begin{tabular}[c]{@{}l@{}}$\mu_S=72.6$, $\sigma_S^{2} = 52.44$\end{tabular} \\ \hline
Number of Buses per route & 1 to 2 \\ \hline
Bus velocity (including stops) & 17.47 to 21.47 km/h \\ \hline
Sensor data generation rate & 10 minutes to 2 hours \\ \hline
Number of buses & 38 \\ \hline
Number of off-route sensors & 100 \\ \hline
Circumference of routes & $\mu=15$ km, $\sigma^{2} = 7$ km \\ \hline
Pedestrian to sensor arrival delay & $\mu=2$ hours, $\sigma^{2} = 30$ mins \\ \hline
$P$ of Pedestrian to Gateway delivery & $\frac{\max(G)}{\mu_S}\approx 0.03$  \\ \hline
Sensor buffer/queue size & $\infty$  \\ \hline
Bus buffer queue/size & $\infty$  \\ \hline
\end{tabular}
\end{table}

\begin{figure}[htbp]
\centering

    \subfloat[]{\includegraphics[width=0.24\textwidth]{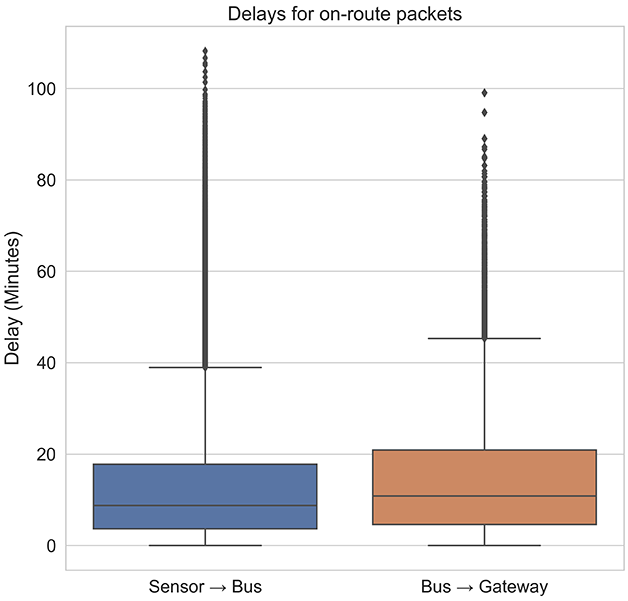}\label{fig_onroute_delay}} 
	\subfloat[]{\includegraphics[width=0.24\textwidth]{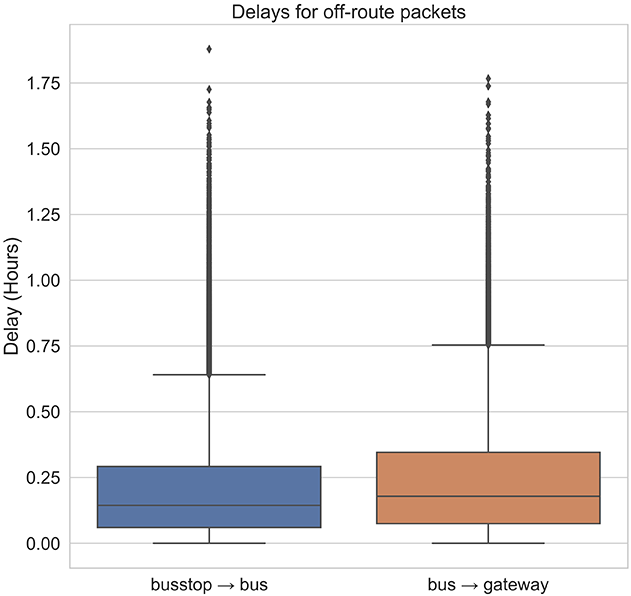}\label{fig_offroute_delay}} 
	
    \subfloat[]{\includegraphics[width=0.24\textwidth]{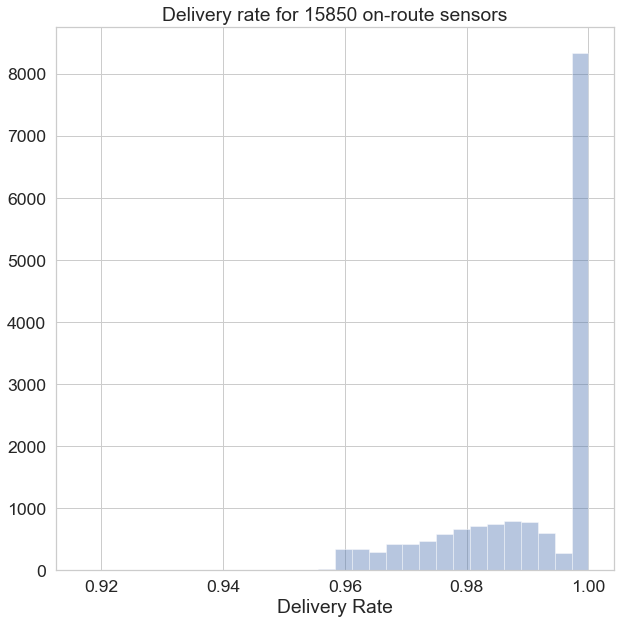}\label{fig_onroute_deliv}} 
	\subfloat[]{\includegraphics[width=0.24\textwidth]{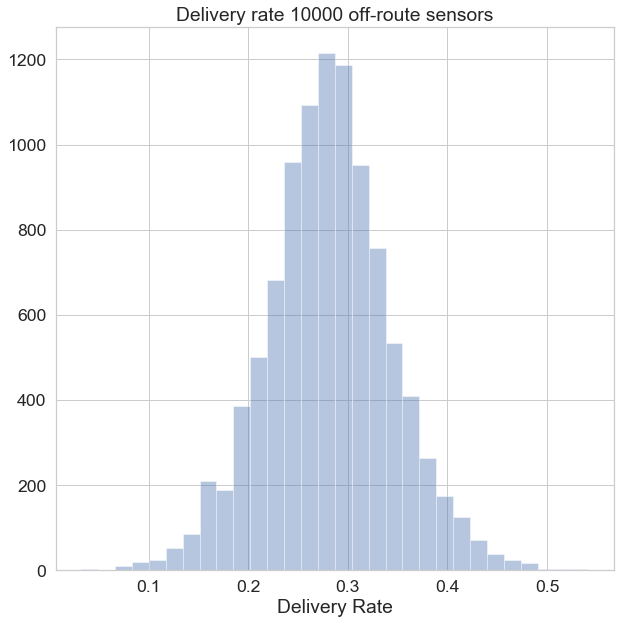}\label{fig_offroute_deliv}} 
\caption{(a) Box plot of delivery delay for on-route sensors, (b) Box plot of delivery delay for off-route sensors, (c) Histogram of delivery rate for on-route sensors, (d) Histogram of delivery rate for off-route sensors.} 
\label{results}
\end{figure}

The results in Figure~\ref{results} are for all messages generated and delivered within the simulation period of $48$ hours for $100$ simulations where the seed was changed from $0-99$. Regarding Figure~\ref{fig_onroute_delay}, half of the on-route sensor data had a delay of $10$ minutes (median) or less before it made it from the sensor onto a bus. Half of the on-route sensor data then had a delay of $15$ minutes (median) or less before it made it from a bus to the gateway.  As expected, the delays of off-route sensor data in Figure~\ref{fig_offroute_delay} was larger as it requires two additional steps (sensor to pedestrian and pedestrian to bus stop) for message delivery. The median time for sensor to pedestrian was $30$ minutes while the time from pedestrian to bus stop was $12.5$ hours.
%
%
In the results, delivery rate is defined as the ratio between messages delivered and the total number of messages generated for each sensor within the simulation duration.
Figure~\ref{fig_onroute_deliv} depicts that all of the on-route sensor data had a delivery rate of $0.96$ or higher. This is due to an assumption in Section~\ref{sec_assumptions} that data can always be transmitted from sensor to bus once contact is made, regardless of the speed of the bus. The aforementioned assumption is built upon another assumption stated in Section~\ref{sec:modeling} that stated data packets are small and can be delivered using Bluetooth 5.
Figure~\ref{fig_offroute_deliv} shows that most of the off-route sensors had a delivery rate of $0.5$ or less. Low delivery rates of off-route sensor data is expected since it is dependent on pedestrians to pick up and deliver the data. Pedestrian mobility is not as predictable as bus mobility, and pedestrian transit ridership rate varies widely as described in the Chapel Hill Transit survey~\cite{CUBIC}.

\section{Future Work}\label{sec:future}

Section~\ref{sec:results} highlighted the poor network latency and low delivery probability of off-route sensors. 
Most sensors in a smart city will not have the benefit of being placed on-route.
Improving off-route sensor performance is the most critical step for DTN's to serve as a viable option as a backbone in smart cities.
%
Policy makers have significant influence over the number of active public transportation vehicles, schedules, and location of sensors and gateways.
They have less control over the number of people who use public transportation, particularly when it comes to pedestrian inter-mobility between public transportation entities.
To drive inter-mobility of pedestrians and non-public transportation vehicles such as ride-share vehicles and taxis to pickup off-route sensor data, the development of effective methods that strategically \textit{incentivize} people is essential for increasing network performance in a low-cost smart city.
Incentives can be generated from needs in network performances and used to drive data requests across the network using a human over-the-loop paradigm~\cite{graham2017formalizing}.
%
%
Classifying data types across the network  to understand expected delays, and tuning the latency restrictions imposed by the infrastructure will drive future research.
%
%
%

%

%

%

\bibliographystyle{IEEEtran}
\bibliography{mendeley_edit,dtnreferences}

\end{document}